    \documentclass[prl,aps]{revtex4}%
\usepackage[dvips]{graphicx}
\usepackage{dcolumn}%
\usepackage{amsmath}%
\usepackage{amsfonts}%
\usepackage{amssymb}
\providecommand{\U}[1]{\protect\rule{.1in}{.1in}}
\newcommand{\be}{\begin{equation}}
\newcommand{\ee}{\end{equation}}
\newcommand{\bea}{\begin{eqnarray}}
\newcommand{\eea}{\end{eqnarray}}
\newcommand{\nn}{ \nonumber}

\begin{document}

\title{On the electron transport in conducting polymer nanofibers}

\author{Natalya A. Zimbovskaya }

\affiliation
{Department of Physics and Electronics, University of Puerto 
Rico-Humacao, CUH Station, Humacao, PR 00791;\\ and
Institute for Functional Nanomaterials, University of Puerto Rico, San Juan, 
PR 00931}

\begin{abstract}
Recent advances in synthesis and electrical characterization of nanofibers and nanotubes made out of various conjugated polymers attract attention of the research community to studies of transport properties of these materials. In this work we present a theoretical analysis of electron transport in polymer nanofibers assuming them to be in conducting state. We treat a conducting polymer as a network of metallic-like grains embedded in poorly conducting environment, which consists of randomly distributed polymeric chains. We analyze the contribution from intergrain electron resonance tunneling via intermediate states localized on the polymeric chains between the grains. Correspondingly, we apply the quantum theory of conduction in mesoscopic systems to analyze this transport mechanism. We show that the contribution of resonance electron tunneling to the intergrain electron transport may be predominating, as follows from experiments on the electrical characterization of single polyaniline nanofibers. 
 We study the effect of temperature on the transport characteristics. We represent the thermal environment as a phonon bath coupled to the intermediate state, which provides electron tunneling between the metallic-like grains. Using the Buttiker model within the scattering matrix formalism combined with the nonequilibrium Green's functions technique, we show that temperature dependencies of both current and conductance associated with the intergrain electron tunneling, differ from those typical for other conduction mechanisms in conducting polymers. Also, we demonstrate that under certain conditions the phonon bath may cause suppression of the original intermediate state accompanied by emergence of new states for electron tunneling. The temperature dependencies of the magnitudes of the peaks in the transmission corresponding to these new states are analyzed.
   \end{abstract}

\pacs{72.15.Gd,71.18.+y}

\date{\today}
\maketitle

\subsection{\normalsize I. Introduction}

During the past decade, transport properties of conducting polymers such as doped polyacetylene and polyaniline-polyethylene oxides, were and still are intensively studied \cite{1,2}.
 These materials are significant mostly due to various possible applications in fabrication of nanodevices. 
Polymer-based devices should have advantages of low cost and flexible, controlled chemistry. Also, there are some unresolved problems concerning the physical nature of charge transfer mechanisms in conducting polymers, which make them interesting subjects for fundamental research.  Chemically doped polymers are known to be very inhomogeneous. In some regions polymer chains are disorderly arranged, forming an amorphous, poorly conducting substance. In other places the chains are ordered and densely packed \cite{3,4}. These regions could behave as metallic-like grains embedded in the disordered environment.
The fraction of metallic-like islands in bulk polymers varies depending on the details of the synthesis process. In practical samples such islands always remain separated by disordered regions and do not have direct contacts. In some cases, electronic  states are delocalized over the grains, and electrons behave as conduction electrons
in conventional metals. In these cases electrons motion inside the grains is diffusive with the diffusion coefficient $D = \frac{1}{3} v^2_F\tau \  (v_F $ is the Fermi velocity, and  $\tau $ is the scattering time). In whole, electron transport in conducting polymers shows both metallic and nonmetallic features, and various transport mechanisms contribute to the resulting pattern.

An important contribution to the conduction in these substances is provided by the phonon-assisted electron  hopping between the conducting islands and/or variable range hopping between localized electronic states. The effect of these transport mechanisms strongly
depends on the intensity of stochastic nuclear motions. The latter increases as 
temperature rises, and this brings a significant enhancement of the corresponding
contributions to the conductivity. The temperature dependence of the ``hopping" 
conductivity $\sigma (T) $ is given by the Mott's expression \cite{5}:
  \be 
 \sigma (T) = \sigma (0) \exp \left[-(T_0/T)^p \right]   \label{1}
  \ee
  where $T_0$ is the characteristic temperature of a particular material, and
the parameter $p$ takes on values $0.25,\,0.33 $ or $0.5 $ depending on the
dimensions of the hopping processes.  
 Also, it was suggested that phonon-assisted transport in low-dimensional structures 
such as nanofibers and nanotubes, may be substantially influenced due to electron
interactions \cite{6,7}. This results in the power-low temperature dependencies
of the conductance $G(T)$ at low values of the bias voltage $V \ (eV < kT,\ k $
being the Boltzmann constant), namely: $G \sim T^\alpha. $ Experimental
data for the conductance of some nanofibers and nanotubes match this power-low
reasonably well, bearing in mind that the value of the exponent $ \alpha$ varies 
within a broad range. For instance, $\alpha $ was reported to accept values 
about $0.35 $ for carbon nanotubes \cite{8}, and $\alpha \sim 2.2 \div 7.2 $ for 
various polyacetylene nanofibers \cite{9,10,11}.
  In general, hopping transport  is very important in disordered 
materials with localized states. For this kind of transport phonons play  part
of a source of electrical conductivity. Accordingly, the hopping contribution
to the conductivity always increases as temperature rises, and more available
phonons appear. When polymers are in the insulating state, the hopping transport
 predominates and determines the temperature dependencies of transport
characteristics.

In conducting state of conducting polymers free charge carriers appear, and their
motion strongly contributes to the conductance. While moving, the charge carriers 
undergo scattering by phonons and impurities. This results in the conductivity
stepping down. Metallic-like features in the temperature dependencies of dc
conductivity of some polymeric materials and carbon nanotubes were repeatedly
reported. For instance, the decrease in the conductivity upon heating was observed
in polyaniline nanofibers in Refs. \cite{12} and \cite{13}.
  However, this electron diffusion is not a unique transport mechanism responsible for the occurrence of metallic-like behavior in the dc conductivity of conducting polymers.
Prigodin and Epstein suggested that  the electron tunneling between the grains through intermediate resonance states on the polymer chains connecting them, strongly contributes to the electron transport \cite{14}. This approach was employed to build up a theory of electron
transport in polyaniline based nanofibers \cite{15} providing
good agreement with the previous transport experiments \cite{16}.
Considering the electron tunneling through the intermediate
state as a mechanism for the intergrain
transport, we see a similarity between the latter and electron transport mechanisms typical for tunnel molecular junctions.  In the case of polymers, metallic-like domains take on part of the leads, and the molecular bridge
in between is simulated by  intermediate sites.  The effect of phonons on this kind of electron transport may be very significant.   These phonons bring an inelastic 
component to the intergrain current and underlie the interplay between the elastic transport by the electron tunneling and the thermally assisted dissipative transport. Also, they may cause some other effects, as was shown while developing the theory of conduction through  molecules \cite{17,18,19,20,21,22,23,24}.

\subsection{\normalsize  II. Electron-resonance tunneling as a transport mechanism in conducting polymers}

Here, we concentrate on the analysis of the electric
current-voltage characteristics and conductance associated with the resonance tunneling transport mechanism. 
 Considering the electron intergrain resonance tunneling, the transmission coefficient  is determined with the probability  of finding the resonance state in between the grains.
 The latter is estimated as $ P \sim \exp(-L/\xi)$  $(L $ is the average distance between the
adjacent grains, and $ \xi $ is the localization length for electrons),
and it takes values much greater than the 
transmission probability for sequental hoppings along the chains, $ P_h 
\sim \exp (-2L/\xi) $ \cite{14}. Nevertheless, the probability for existence of a resonance 
state at a certain chain is rather low, so only a few out of the whole set
of  chains connecting two grains are participating in the 
intergrain electron transport. Therefore, one could assume that any two metallic
domains are connected by a single chain providing an intermediate state for
the resonance tunneling. All remaining chains can be neglected for they
poorly contribute to the transport compared to the resonance chain.
Within this approximation the ``bridge" linking two islands is reduced to a single electron state. Realistic polymer nanofibers have diameters within the range $ 20\div 100$nm,
and lengths of the order of a few microns. This is much greater than a
typical size of both metallic-like grains and intergrain separations, which 
take on values $ \sim 5\div 10$nm (see e.g. Refs. \cite{16} and \cite{25}). Therefore, we may 
treat a nanofiber as a set of working channels connected in parallel, any 
single channel being a sequence of grains connected with the resonance polymer 
chains. The net current in the fiber is the sum of currents flowing in these 
channels, and the voltage $ V$ applied across the whole fiber is distributed 
among sequental pairs of grains along a single channel. So, the voltage 
$ \Delta V$ applied across two adjacent grains could be roughly estimated as 
$ \Delta V \sim V{L/L_0} $ where $ L $ is the average separation between the 
grains, and $ L_0 $ is the fiber length. In practical fibers the ratio $ \Delta V/V $ may take on values of the order of $  10^{-2} \div 10^{-3} . $ 

 In further current calculations we treat the grains as free electron reservoirs in thermal equilibrium. This
assumption is justified when the intermediate state (the bridge) is weakly
coupled to the leads, and conduction is much smaller than the  quantum 
conductance $G_0 = 2e^2/h\ (e,h $ are the electron charge, and the Planck
constant, respectively). Due to the low probabilities for the resonance
tunneling between the metallic islands in  conducting polymers, the above 
assumption may be considered as a reasonable one. So, we can employ the 
well-known expression for the electron current through the molecular junction \cite{26},
and we write:
  \be %
 I = \frac{2en}{h} \int_{-\infty}^\infty dE T(E) [f_1(E) - f_2(E)].   \label{2}
   \ee  
  Here, $ n $ is the number of the working channels in the fiber, $f_{1,2} (E)$ 
are Fermi functions taken with the different contact chemical potentials 
$ \mu_{1,2} $ for the grains. The chemical potentials differ due to the  
bias voltage $ \Delta V $ applied across the grains:
  \be 
 \mu_1 = E_F + (1 - \eta) e \Delta V; \qquad
 \mu_2 = E_F - \eta e \Delta V.                     \label{3}
  \ee
   In these expressions (\ref{2}), (\ref{3}) the parameter $ \eta $ characterizes how the voltage $ \Delta V $ is divided between the grains, $ E_F $ is the equilibrium Fermi energy of the system 
including the pair of grains and the resonance chain in between, and $ T (E) $ 
is the electron transmission function.

The general approach to the electron transport  studies 
in the presence of dissipation is the reduced dynamics density-matrix
formalism (see, e.g., Refs. \cite{27} and \cite{28}). This microscopic
computational approach has the advantages of being capable of 
providing the detailed dynamics information. However, this information is usually more redundant
than necessary, as far as standard transport experiments in conducting
polymer nanofibers are concerned. There exists  an alternative approach using the scattering matrix formalism and the phenomenological Buttiker dephasing model \cite{29}. Adopting this phenomenological model we are able to analytically study the problem, and
the results agree with those obtained by means of more sophisticated computational methods, as was demonstrated in the earlier works \cite{30}.

Within the Buttiker model we treat the intergrain electron transport as a multichannel scattering problem. In the considered case  the ``bridge" between two adjacent grains inserts a single electron state. Therefore, an electron could be injected into the system (including two metallic-like domains and the intermediate ``bridge" in between) and/or leave from there via four channels presented in the Fig. 1. The electron transport is  a combination
of tunneling through two barriers (the first one separates the left metallic
island from the intermediate state  and
the second separates this state from the right island, supposing the transport
from the left to the right).  Inelastic effects are accounted for by means of a dissipative electron reservoir attached to the bridge site. The dissipation strength is characterized by a phenomenological parameter $ \epsilon ,$ which can take values within the range $[0,1]. $ When $ \epsilon = 0 $ the reservoir is detached from the bridge, which corresponds to the elastic and coherent electron transport. The greater is $ \epsilon $ value the stronger is the dissipation. 
  In the Fig. 1 the barriers are represented by the squares, and the 
triangle in between imitates a scatterer coupling the bridge to a dissipative 
electron reservoir.

\begin{figure}[t] 
\begin{center}
\includegraphics[width=4cm,height=10cm,angle=-90]{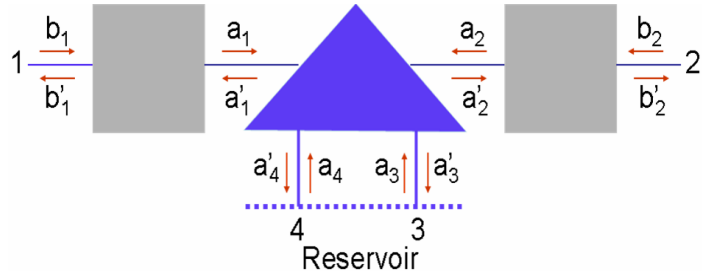}
\caption{ Schematic drawing illustrating the intergrain electron transport
in the presence of dissipation \cite{29}. Rectangles correspond to the barriers separating the adjacent metallic-like islands from the intermediate state (the bridge), and the triange stands for a scatterer attached to the bridge.
}  
\label{rateI}
\end{center}
\end{figure}

 Incoming particle fluxes 
$(J_i)$ are related to those outgoing from the system $(J_j')$ by means of the
transmission matrix $T$ \cite{29,30}:
  \be 
  J_j' = \sum_i T_{ji} J_i, \qquad 1\leq i,  j \leq 4.     \label{4}
  \ee
  Off-diagonal matrix elements $T_{ji} (E) $ are probabilities for an electron
to be transmitted from the channel $ i $ to the channel $j, $ whereas diagonal
matrix elements $T_{ii}(E) $ are probabilities for its reflection back to the
channel $i $. To provide charge conservation, the net particle flux
in the channels connecting the system with the reservoir must be zero. So we
have:
  \be 
J_3 + J_4 - J_3' - J_4' = 0.            \label{5}
  \ee 
 The transmission function $ T(E) $ relates the particle flux outgoing from 
the channel $2$ to the flux incoming to the channel $1,$ namely:
  \be 
J_2' = T(E) J_1.                \label{6}
 \ee
   Using Eqs. ({\ref{4}) and (\ref{5}) we can express the transmission function in
terms of the matrix elements $T_{ji}. $The latter are related to matrix elements of the scattering matrix $ S, $ which expresses the 
outgoing wave amplitudes $b_1',b_2',a_3', a_4'$  as linear combinations of  the incident ones 
$b_1,b_2,a_3, a_4:\ T_{ij} = |S_{ij}|^2. $ In the considered case 
of a single site bridge the $S$ matrix takes the form \cite{15,30}:
  \be 
 S = Z^{-1}\left(
   \begin{array}{cccc}
r_1 + \alpha^2 r_2 & \alpha t_1 t_2 & \beta t_1 & \alpha\beta t_1 r_2 
   \\
\alpha  t_1 t_2 & r_2 +\alpha^2 r_1 & \alpha\beta r_1 t_2 &  \beta t_2
   \\
 \beta t_1 & \alpha \beta r_1 t_2 & \beta^2 r_1 & \alpha r_1 r_2 - \alpha
   \\
\alpha \beta t_1 r_2 & \beta t_2 & \alpha r_1 r_2 - \alpha & \beta^2 r_2
    \end{array} \right) .                   
                            \label{7}     \ee
  where $Z = 1 - \alpha^2 r_1 r_2,\ \alpha = \sqrt{1- \epsilon}, \ \beta =
\sqrt\epsilon,$ $r_{1,2}$ and $t_{1,2} $ are the transmission an
reflection coefficients for the barriers $(|t_{1,2}|^2 + |r_{1,2}|^2 = 1).$

When the bridge is detached from the dissipative reservoir $ T(E) = |S_{12}|^2.$
On the other hand, in this case we can employ a simple analytical expression 
for the electron transmission function \cite{31}:
  \be 
  T(E) = 4 \Delta_1(E) \Delta_2 (E) |G(E)|^2,      \label{8}
  \ee
 where $\Delta_{1,2} (E) = - \mbox{Im} \Sigma_{1,2} (E). $ In this expression,
self-energy terms $ \Sigma_{1,2} $ appear due to the coupling of the metallic-like
grains to the intermediate state (the bridge).
  The retarded Green's function for a single-site bridge could be approximated as follows:
  \be 
  G(E) = \frac{1}{E -E_1 +i\Gamma}         \label{9}
  \ee
  where $E_1 $ is the site energy. The width of the resonance level between
the grains is determined by the parameter $ \Gamma = \Delta_1 + \Delta_2 + 
\Gamma_{en}\ (\Gamma_{en}$ describes the effect of energy dissipation). Further we consider dissipative effects originating from electron-phonon interactions, so, $\Gamma_{en} $ is identified with $ \Gamma_{ph}. $

Equating the expression (\ref{8}) and $ |S_{12}|^2 $ we arrive at the following
expressions for the tunneling parameters $ \delta_{1,2} (E) :$
   \be 
 \delta_{1,2}(E)\equiv t_{1,2}^2 = 
\frac{2 \Delta_{1,2}}{\sqrt{(E- E_1)^2 + \Gamma^2}} . \label{10}
  \ee

Using this result we easily derive the general expression for the electron
transmission function:
    \be 
  T (E) = \frac{g(E)(1+ \alpha^2) [g(E)(1+ \alpha^2) + 1 
- \alpha^2]}{[g(E)(1 - \alpha^2) + 1 + \alpha^2]^2}         \label{11}
   \ee
  where:
    \be 
  g(E) = 
2 \sqrt{\frac{\Delta_1\Delta_2}{(E - E_1)^2 + \Gamma^2}}.            \label{12}
  \ee

To simplify further analysis  we approximate the self-energy terms $ \Delta_{1,2} $ as constants: $ \Delta_{1,2} \approx W_{1,2}^2/ \sigma_{1,2}. $ Here, $ W_{1,2} $ are coupling strengths characterizing the coupling of the grains to the bridge and $ \gamma_{1,2} $ characterize interatomic couplings inside the grains (leads). Simulating the leads by semiinfinite chains of identical sites, as was first  suggested by D'Amato and Pastawski \cite{32}, one may treat the parameters $ \gamma_{1,2} $ as coupling strengths between the nearest sites in these chains. 
  Then the elastic electron transmission $(\epsilon = 0)$ shows a sharp
peak at the energy $ E_1 $  (see Fig. 2a), which gives rise to a
steplike form of the volt-ampere curve presented in Fig. 2b.
When the  reservoir is attached to the electronic
bridge $(\epsilon \neq 0)$, the peak in the transmission is eroded. The
greater is the value of the parameter $\epsilon $, the stronger
is the erosion. When $\epsilon $ takes on the value of $ 0.7 $ the peak in the
electron transmission function is completely washed out as
well as the steplike shape of the $ I-V $ curve. The latter becomes
linear, corroborating the well-known Ohmic law for the sequential hopping mechanism.  So, we see that the electron transmission is affected by  stochastic nuclear motions in the environment of the resonance state.   When the dissipation is strong (e.g. within the strong thermal coupling limit), the inelastic (hopping)  contribution to the intergrain current predominates, replacing the coherent elastic tunneling. Typically, at room 
temperatures the intergrain electron transport in conducting polymers occurs
within an intermediate regime, when both elastic and inelastic contributions
to the electron transmission are manifested.

\begin{figure}[t] 
\begin{center}
\includegraphics[width=6cm,height=11.8cm,angle=-90]{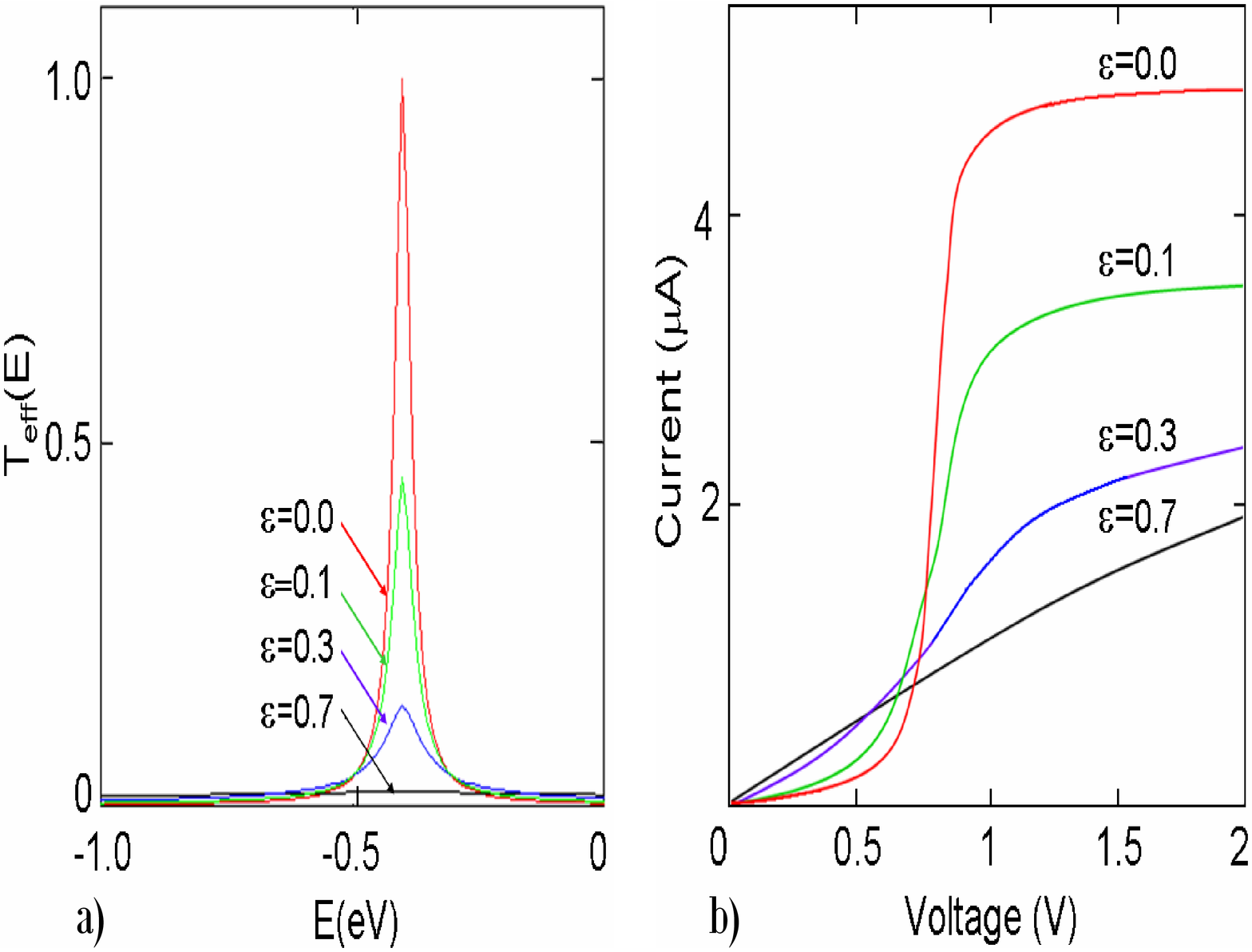}
\includegraphics[width=6cm,height=11.8cm,angle=-90]{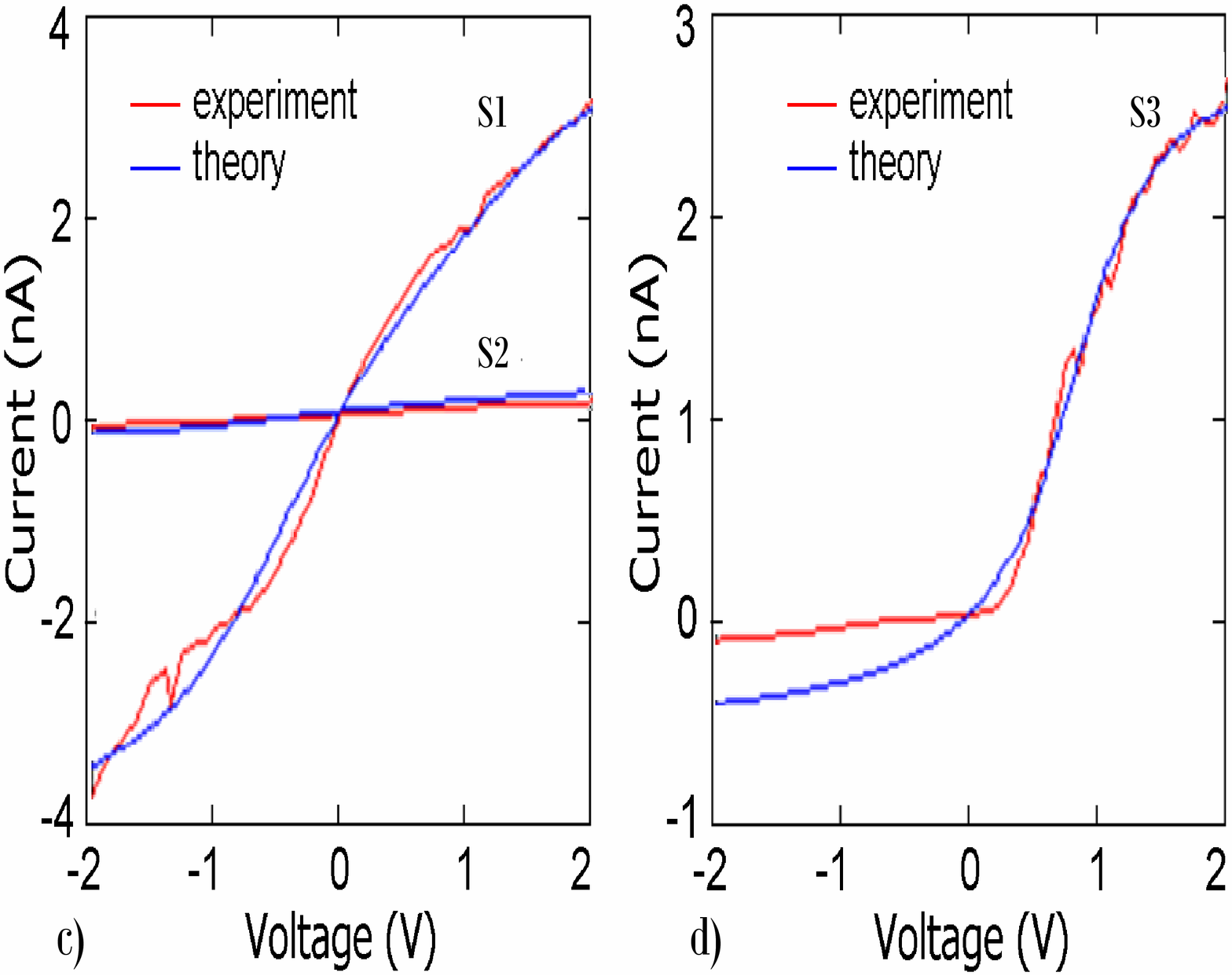}
\caption{Dissipation effect on the effective transmission function (a) and the current voltage characteristics (b).  The curves are plotted at $ T= 70 K,\ \Delta_1 = \Delta_2  = 10 meV,\ \epsilon = 0.0, \ 0.1, \ 0.3 ,\ 0.7 $ from the top to the bottom. Calculated  (dash-dot lines) and experimental (solid lines) current (nA) -- voltage (V) characteristics for PANi nanofibers: (c) samples 1 (S1), and 2 (S2), and (d) sample 3 (S3). } 
\label{rateI}
\end{center}
\end{figure}

The described approach was succesfully employed to analyze experimental results on the electrical characterization of polyaniline-polyethylene oxides (PANi) nanofibers reported by Zhou et al \cite{16}, and Pinto et al \cite{33}. In these experiments  single-fiber electrical characterization was carried out at  $T \sim 300 K.$ The experiments revealed that current-voltage curves for conducting nanofibers were non-Ohmic.  
  Conducting samples included a 70-nm-diameter nanofiber (sample 1), a pair
of 18-nm- and 25-nm-diameter fibers connected in parallel
(sample 2) and a fiber whose diameter was steeply reduced in the middle from 70 nm to 20 nm (sample 3). The latter did exhibit a very asymmetric rectifying current-volltage characteristic. The expression (\ref{2}) was employed to compute the current originating from electron resonance tunneling in the nanofibers \cite{15}. To evaluate the number of working channels in a nanofiber it was assumed that  for certain doping and crystallinity rates the number of channels is proportional to the fiber cross-sectional area. Also, it was taken into account that the contact of the fiber surfaces with atmospheric gases reduced the number of working channels.
    Accepting the value 5 nm for both
average grain size and intergrain distance  we estimated  that the 70 nm fiber  could include about $ 30 - 40 $ conducting
channels, and we could not expect more than two
working channels in the pair of fibers (sample 2). This gives
the ratio of conductions $ \sim 15-20, $ which is greater than the
ratio of cross-sectional areas of the samples $A_1 /A_2 \approx 5$. The
difference originates from the stronger dedoping of thinner fibers. Calculating the current, appropriate values for $ W_{1,2} $ providing the best match between  calculated and experimental $ I-V $ curves were found applying the least-squares procedure. We estimated values of the coupling strengths $ W_{1,2}$  in PANi fibers  as $ 2.0-3.5  meV. $  As shown in the Fig. 2, computed  volt-ampere characteristics
match the curves obtained in the experiments reasonably well.   
The  agreement between the presented theoretical
results and experimental evidence proves that electron
tunneling between metallic-like grains through intermediate
states at the resonance chains can really play a significant part in
the transport in conducting polymers provided that they are
in the metallic state. 

However, we remark that the good fitting between the theory and the experiment demonstrated in the work \cite{15} was achieved assuming that dissipation was very weak $ (\alpha \approx 0.05).$ This assumption could hardly be justified while one is observing electron transport in nanofibers at room temperature. So, the phenomenological approach employed in this section  has some significant shortcomings. Its main disadvantage is that the dissipative effects are described in terms of a phenomenological  parameter $\epsilon, $ whose dependence of  characteristic factors affecting the transport (such as  temperature, electron-phonon coupling strength and some others) remains unclear.  It is necessary to modify the Buttiker's model  to elucidate the relation of the parameter $\epsilon $ to the relevant energies characterizing  electron transport in the considered systems.
 To obtain the desired expression for $\epsilon $, we compare our results with those  presented in Ref. \cite{17}. In that work the inelastic correction $\delta I $ to the coherent tunnel current via a single-site bridge is calculated using the nonequilibrium Green's functions formalism. The relevant result  is derived in the
limit of weak electron-phonon interaction when $ \Gamma_{ph}  \ll \Delta_{1,2}. $
 It is natural to assume that the dissipation strength  $\epsilon $ is small within this limit, so we expand our expression for $ T(E) $ in powers of $\epsilon. $ Keeping  two first
terms in this expansion,  and assuming that $ \Delta_1 = \Delta_2 \equiv \Delta $ we obtain:
   \be
 T (E) \approx   g^2 (E) \Big(1 + \frac{\epsilon}{2} [1 - 2g^2(E)]\Big).                                                         \label{13}                                    \ee
 We employ this approximation to calculate the current through the bridge, and we arrive at the following expression  for $\delta I$:
  \begin{align}
 \delta I = & \frac{e}{\hbar} \int_{-\infty}^\infty dE \frac{\epsilon \Delta}{\Gamma + \Gamma_{ph}} \rho_{el}(E) \frac{(E -  E_1)^2 - 4 \Delta^2}{(E - E_1)^2 + 4 \Delta^2}  
\nn\\    & \times\big [f_1(E ) - f_2(E) \big ]                  .       \label{14}
 \end{align}   
Here, $ \rho_{el} (E)$ is the electron density of states at the bridge:

  \be
 \rho_{el}(E) = - \frac{1}{\pi}\mbox{Im} G(E).     \label{15}
 \ee

Comparing the expression (\ref{14}) with the corresponding
result of Ref. \cite{17}, we find  that these two are consistent, and we get \cite{34}:

   \be
 \epsilon = \frac{\Gamma_{ph}}{\Gamma + \Gamma_{ph}} .       \label{16}
   \ee
When $ \Delta \gg \Gamma_{ph} \ (\epsilon \ll 1) $ the bridge coupling to the dissipative reservoir is weak, and the elastic electron tunneling predominates.The opposite limit $ \Delta \ll \Gamma_{ph} \ (\epsilon \sim 1) $ corresponds to the completely incoherent phonon-assisted electron transport. 

\subsection{\normalsize  III. Temperature Dependencies of electron transport characteristics in conducting polymer nanofibers}

Now, we turn to studies of  temperature dependencies of the electron current and conductance associated with the intergrain electron tunneling.  It is known that various conduction mechanisms may simultaneously contribute to the charge transport in conducting polymers, and their relative effects could significantly differ depending on the specifics of synthesis and processing of polymeric materials. The temperature dependencies of the resulting transport characteristics may help to identify the predominating transport mechanism for a paricular sample under particular conditions.
  The issue is of a significant  importance  because the relevant transport 
experiments are often implemented at room temperature, 
so that the influence of phonons cannot be disregarded.
Therefore, we study the effect of temperature (stochastic nuclear motions)
on the  resonance electron tunneling between metallic-like grains (islands)
in polymer nanofibers. In these studies we consider the dissipative reservoir attached to the ``bridge" site as a phonon bath, and 
 we assume that the phonon bath is characterized by the 
continuous spectral density $J (\omega)$ of the form \cite{35}:
  \be 
J(\omega) = J_0 \frac{\omega}{\omega_c} \exp \left(-\frac{\omega}{\omega_c}
\right )             \label{17}
  \ee
  where $ J_0 $ describes the electron-phonon coupling strength, and 
$ \omega_c $ is the cut-off frequency of the bath, which determines the thermal 
relaxation rate of the latter.   The expression for $\Gamma_{ph} (E)$ was derived  in  earlier works \cite{17,36,37} using nonequilibrium Green's functions approach. Using these results and the expression (\ref{17}), we may present $ \Gamma_{ph} (E) $ as follows:
  \begin{align}
 \Gamma_{ph} =& 2\pi J_0 \int d\omega 
\frac{\omega}{\omega_c} \exp \left(-\frac{\omega}{\omega_c}
\right )
    \nn\\ & \times
\big \{N(\omega) [\rho_{el}(E-\hbar\omega)
 +\rho_{el}(E+\hbar\omega)]
 \nn\\ &
  +  [1-n(E-\hbar\omega) ]\rho_{el} (E-\hbar\omega)
   \nn\\& 
+  n(E + \hbar\omega) \rho_{el} (E+\hbar\omega) \big\}.      \label{18}
 \end{align}
 Here,
    \be 
n(E) = \frac{\Delta_1 f_1(E) + \Delta_2 f_2 (E)}{\Delta_1 + \Delta_2},
             \label{19}                 \ee
  and      $ N(\omega) $ is the Bose-Einstein distribution  
function for the phonons at the temperature $T. $ The asymptotic expression 
for the self-energy term $ \Gamma_{ph} $ depends on the relation between two 
characteristic energies, namely: $ \hbar \omega_c $ and $ kT \ (k$ is the 
Boltzmann constant). At moderately low or room temperatures  $kT \sim 10\div 30 meV.$ This is significantly greater than typical 
values of $ \hbar \omega_c \ (\hbar\omega_c \sim 1 meV $ \cite{17}). Therefore, 
in further calculations we assume $ \hbar \omega_c\ll kT. $ Under this 
assumption, the main contribution to the integral over $ \omega $ in the Eq. 
(\ref{18}) originates from the region where $ \omega\ll \omega_c \ll kT/\hbar, $ 
and we can use the following approximation \cite{38}:
  \be 
 \Gamma_{ph} (E) = \frac{2 \Gamma \Lambda (J_0,\omega_c, T)}{(E- E_1)^2 + 
\Gamma^2}.          \label{20}
      \ee
  Here, $\Gamma = \Delta_1 + \Delta_2 + \Gamma_{ph}; $
       \be 
\Lambda = \frac{4 J_0}{\hbar \omega_c} (kT)^2 \zeta 
\left(2; \frac{kT}{\hbar\omega_c} + 1\right)           \label{21}
  \ee 
 where $\zeta (2; kT/\hbar\omega_c + 1) $ is the Riemann $ \zeta $ function: 
  \be 
\zeta = (2;kT/\hbar\omega_c + 1) = \sum_{n=1}^\infty \frac{1}{(n+ kT/\hbar
\omega_c)^2}.             \label{22}
  \ee
  Under $ \hbar\omega_c \ll kT, $ we may apply the estimation $ \Lambda 
\approx 4k TJ_0. $

Solving the equation (\ref{20}) we obtain a reasonable asymptotic expression for 
$ \Gamma_{ph}: $ 
   \be
\Gamma_{ph} = \frac{\Delta_1 + \Delta_2}{2} \frac{\rho^2(1 + \sqrt{1 + 
\rho^2})}{4\big(\frac{E - E_1}{\Delta_1 + \Delta_2}\big)^2 + (1 + \sqrt{1 + 
\rho^2})^2} .                \label{23}
  \ee
  where $ \rho^2 = 8 \Lambda/(\Delta_1 + \Delta_2)^2. $
 Substituting this expression into Eq. (\ref{16}) we arrive at the result for the 
dissipation strength $ \epsilon :$
  \be 
 \epsilon = \frac{1}{2}\frac{\rho^2(1 + \sqrt{1 + \rho^2})}{4\big
(\frac{E 
- E_1}{\Delta_1 + \Delta_2}\big)^2 +\frac{1}{2} (1 + \sqrt{1 + \rho^2})^3} 
       \label{24} \ee
  This expression shows how the $ \epsilon $ depends on the temperature 
$ T ,$ the electron-phonon coupling strength $ J_0, $  and the energy $ E. $ In 
particular, it follows from the Eq. (\ref{24}) that $ \epsilon $ reachs its maximum 
at $ E = E_1, $ and the peak value of this parameter is given by:
  \be 
\epsilon_{max} = \frac{\sqrt{1 + \rho^2}-1}{\sqrt{1 + \rho^2}+1}.  \label{25}
  \ee 
 
\begin{figure}[t]  
\begin{center}
\includegraphics[width=11.8cm,height=6cm]{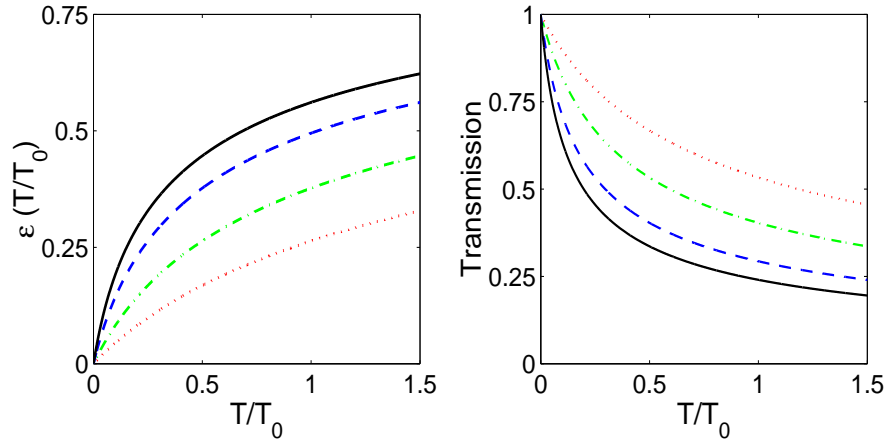}
\caption{ Temperature  dependencies of the maximum dissipation parameter $\epsilon$ 
(left panel) and the electron transmission (right panel).
 The curves are plotted at $ T_0 = 50K,\ \Delta_1 = \Delta_2 = 4 meV,\ 
E = E_1 = 0,\  J_0 = 9.0 meV$ (solid line), $6.0 meV $ (dashed line), $3 meV $ 
(dash-dotted line), $1.5  meV$ (dotted line). 
}  
\label{rateI}
\end{center}
\end{figure}

 The maximum value of the dissipative strength is determined with two parameters, 
namely, $ T $ and $ J_0. $ As illustrated in the Fig. 3, $ \epsilon_{\max}$ increases when the temperature rises, and it takes on  greater values when the electron-phonon interaction is getting stronger. This result has a clear physical sense. Also, as follows from the
Eq. (\ref{24}), the dissipation parameter  exhibit a peak at $ E = E_1 $ whose shape is determined by the product $ kTJ_0 .$ When either $ J_0$ or $ T $ or both enhance, the peak becomes higher and its width increases. The manifested energy dependence of the dissipation strength allows us to resolve the above mentioned  difficulty occurring when the inelastic contribution to the electron transmission function is estimated using the simplified 
approximation of the parameter $ \epsilon $ as a constant. When the energy dependence of $\epsilon$ is accounted for, the peak in the electron transmission at $E=E_1$ may be still distinguishable when $ \epsilon_{\max} $ takes on values as big as 0.5.

\begin{figure}[t]  
\begin{center}
\includegraphics[width=12cm,height=11.8cm,angle=-90]{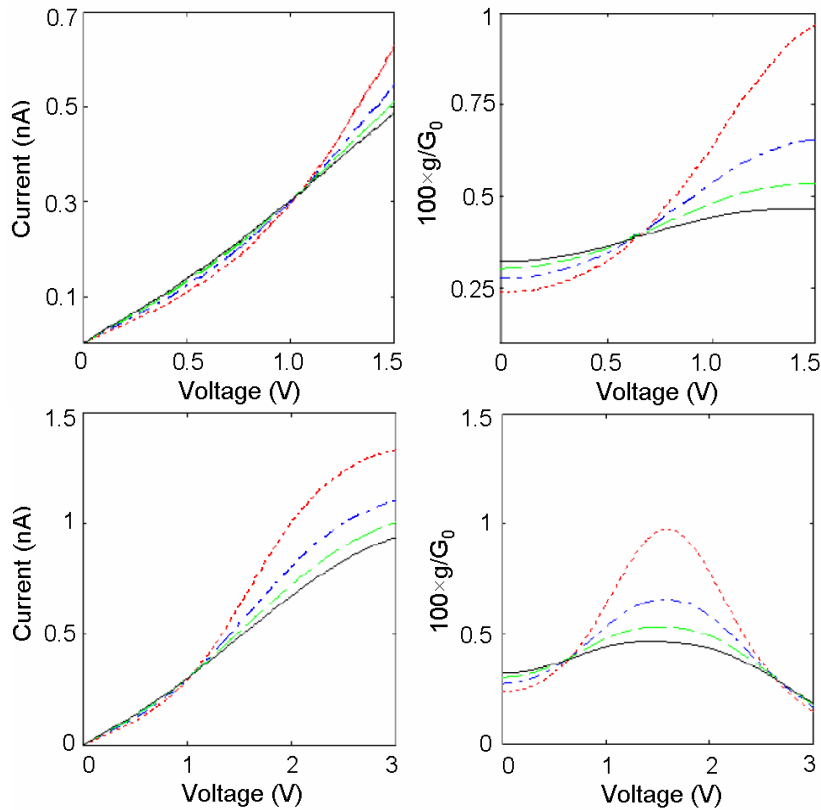}
\caption{ Current (left panels) and conductance (right panels) versus voltage.
The curves are plotted at $ T_0= 50K,\ T = 30 K,\ E_1 = 40 meV,\ 
\Delta_1 = \Delta_2= 4meV,\ n=1, \ \Delta V/V=0.005, $ 
assuming $J_0 = 9meV $ (solid line), $J_0 = 6meV $ (dashed line),
$ J_0= 3meV $ (dash-dotted line), and $J_0 = 1.5 meV $ (dotted line).
$G_0 = 2e^2/h$ is the quantum conductance.
}  
\label{rateI}
\end{center}
\end{figure}

  The obtained result enables us to analyze the temperature dependencies of the
electric current and conductance of the doped polymer fibers assuming that the
resonance tunneling predominates in the intergrain electron transport in the 
absence of phonons.
Current-voltage characteristics and voltage dependencies of the conductance
$ G = dI/dV $ computed using the expressions (\ref{2}), (\ref{11}), (\ref{24}) are 
presented in the Fig. 4. We see that   at low values of the applied voltage the electron-phonon coupling brings an enhancement in both current and conductance, as shown in the top panels of the Fig. 4. The effect becomes reversed as the voltage grows above a certain value (see Fig. 4, the bottom panels).  This happens because the phonon induced broadening of the intermediate energy level (the bridge)  assists the electron transport at small bias voltage. As the voltage rises, this effect is surpassed by the scattering effect of
phonons which resists the electron transport.
 When  the electron-phonon coupling 
strengthens, the I-V curves lose their specific shape typical for the elastic
tunneling through the intermediate state. They become closer to
straight lines corresponding to the Ohmic law. At the same time the maximum in 
the conductance originating from the intergrain tunneling gets eroded.
  These are the obvious results discussed in some earlier
works (see e.g. Ref. \cite{30}). The relative strength of the electron-phonon
interaction is determined by the ratio of the electron-phonon coupling constant
$J_0 $ and the self-energy terms describing the coupling of the intermediate 
state (bridge) to the leads $\Delta_{1,2}.$ The effect of phonons on the
electron transport becomes significant when $J_0> \Delta_{1,2}.$
   Otherwise, the coherent tunneling between the metallic-like islands prevails in the intergrain electron transport, and the influence of thermal phonon bath is weak. Again, we may remark that $ J_0$ and $ T $ are combined as $ kTJ_0 $  in the expression (\ref{24}).  Therefore, an increase in temperature at a fixed electron-phonon coupling strength enhances  the inelastic contribution to the current in the same way as the previously discussed increase in the electron-phonon coupling. 

\begin{figure}[t]  
\begin{center}
\includegraphics[width=6cm,height=16.2cm,angle=-90]{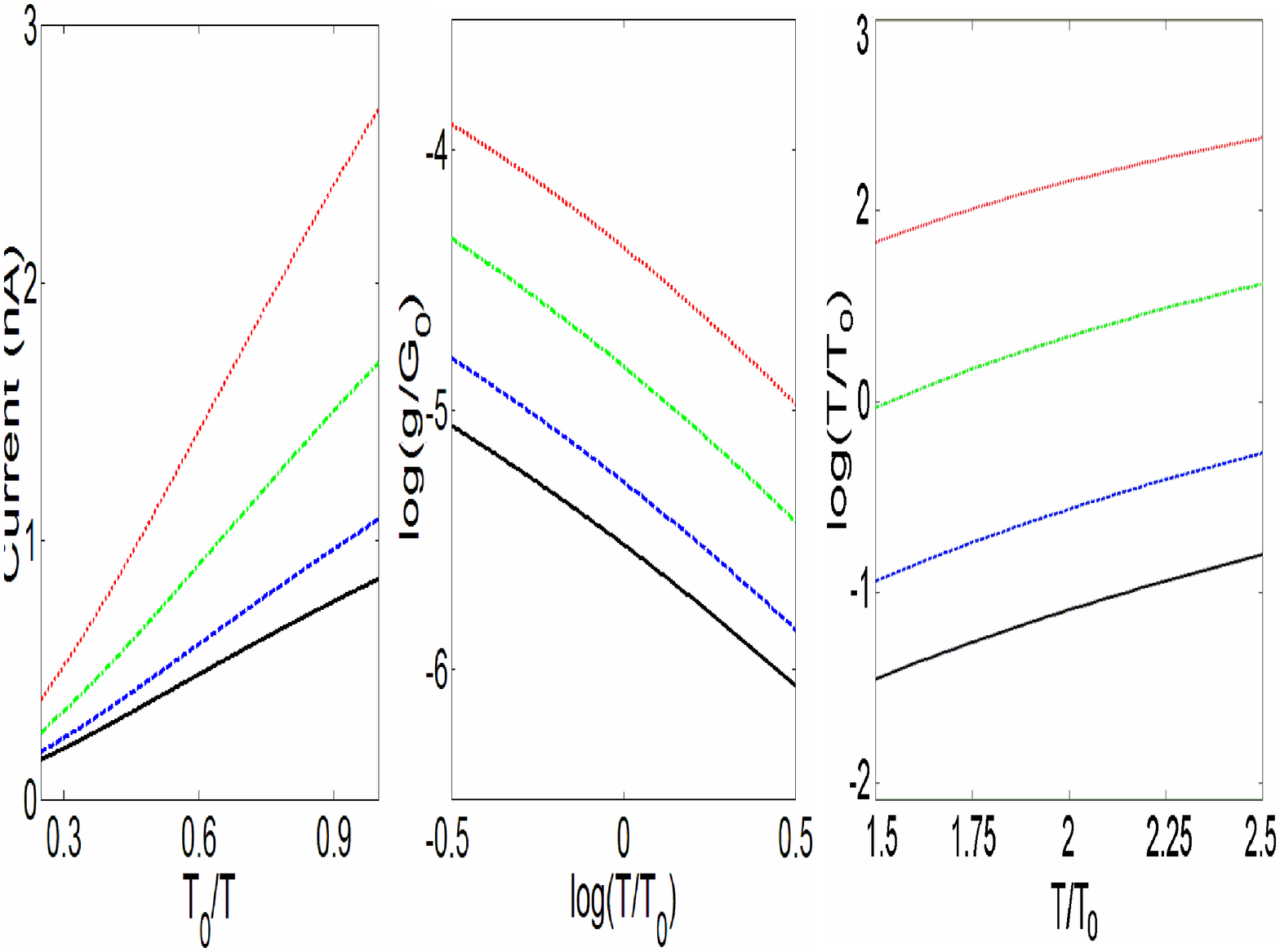}
\caption{ Temperature dependence of current (left panel) and conductance (right panel) at low voltage bias $(V = 0.3 V).$  The curves are plotted at $ T_0 = 50 K, $ 
 $J_0 = 9meV $ (solid line), $J_0 = 6meV $ (dashed line),
$ J_0= 3meV $ (dash-dotted line), and $J_0 = 1.5 meV $ (dotted line). The  values 
of remaining parameters are the same as in the Fig. 4. The current $I_0 $
is computed using $J_0 = 3 meV;\ T_0/T = 1.5.$
}  
\label{rateI}
\end{center}
\end{figure}

Now, we consider temperature dependencies of the electric current and conductance resulting from the intergrain electron tunneling via the intermediate localized state. These dependencies are shown in the Fig. 5. The curves in the figure are plotted at low bias voltage $(V= 0.3 V,\ \Delta V/V = 0.005)$ and $T_0 = 50 K,$ so $e\Delta V < kT.$  This regime is chosen to compare the obtained temperature dependencies with those typical for the phonon assisted hopping transport discussed. We see that the temperature dependence of the tunnel current shown in the left panel of Fig. 5 crucially disagrees with the Mott's expression (\ref{1}). The tunnel current decreases as temperatute rises being proportional to $(T_0/T)^\beta, $  and the exponent $\beta $ takes on values close to unity.

Already it was mentioned that the drop in the conductivity upon heating a sample was observed in polymers and carbon nanotubes. However, such metallic-like behavior could originate from various dc transport mechanisms. Correspondingly, the specific features of temperature dependencies of the conductivity and/or current vary depending on the responsible conduction
mechanism \cite{39}. The particular temperature dependence of the electron tunneling current  shown in Fig. 5 differs from those occurring due to other transport mechanisms. Such dependence was observed in the experiment on the electron transport in a single low-defect-content carbon nanotube rope, whose metallic-like conductivity was manifested within a wide temperature range $(T\sim 35\div 300K),$ as reported by Fisher et al \cite{40}. The conductivity temperature dependence observed in this work could be approximated as $\sigma (T)/ \sigma(300) \sim a+ bT_0/T $ where $A,b$ are dimensionless constants. The approximation includes the temperature independent term, which corresponds to the Drude conductivity. The second term is inversely proportional to the temperature in agreement with 
the  results for the current shown in the Fig. 5 (left panel). It is also likely that a similar
approximation may be adopted to describe  experimental data obtained for chlorate-doped polyacetylene samples at the temperatures below $100 K$ \cite{41}. In both cases we may attribute the contribution proportional to $ 1/T$ to the resonance electron tunneling transport between metallic-like islands.

The conductance due to the electron intergrain tunneling reduces when  the temperature increases, as shown in the right panel of the Fig. 5. Irrespective of the electron-phonon coupling strength we may approximate the conductance by a power law $G\sim T^\alpha $ where $\alpha $ takes on values close to $ -1. $ This agrees with the results for the current.  At higher bias voltage the temperature dependence of the current  changes, as shown in  the Fig. 6. The curves shown in this panel could be approximated as $\log (I/I_0) \sim c + d T_0/T;\ c,d$ being dimensionless constants. This resembles typical temperature dependencies of the tunneling current in quasi-one-dimensional metals which were predicted for conducting polymers being in a metal state (see e.g. Ref. \cite{2}).


\subsection{\normalsize  IV. Effect  of phonon induced electron states on the transport properties of conducting polymer fibers}

 Studies of dissipative effects in the intergrain electron transport in conducting polymers may not be restricted with the plain assumption of direct coupling of the bridge site to the phonon bath. Other scenarios can occur. In particular, analyzing  electron transport in polymers, as well as in molecules, one must keep in mind that besides  the bridge sites there always exist other nearby sites with close energies. In some cases the presence of such sites may strongly influence the effects of stochastic nuclear motions on the characteristics of electron transport. This may happen when the nearby sites somehow ``screen" the bridge sites from direct interactions with the phonon bath. Here,  we elucidate some effects which could appear in the electron transport in conducting polymer fibers in the case of such indirect coupling of the bridge state to the phonon bath.

We mimic the effects of the environment by assuming that the side chain is attached to the bridge, and this chain is affected by phonons. This model resembles those used to analyze electron transport through macromolecules \cite{19}.  The side chain is introduced  to screen the resonance state making it more stable against the effect of
phonons. We assume that electrons cannot hop along the side chain, so it may be reduced to a single site attached to the resonance site (the bridge).

 Within the adopted model the retarded Green's function for the bridge acquires the form \cite{19,42}:
    \be 
  G^{-1} (E) = E - E_1 - \Delta_1  - \Delta_2 - w^2 P(E).    \label{26}
    \ee
 The first four terms in this expression represent the inversed Green's function for the resonance site coupled to the two grains,  and the factor $w $ is the hopping integral between the bridge and the attached side chain.  The  term $P(E) $ represents the effect of the phonons and has the form:
 \begin{align}
  P(E) =&\!\! -i \int_0^\infty dt \exp \big[it (E - \tilde E_1 + \delta + i 0^+)\big]
   \nn \\ & \times \!\!
\Big \{(1 - f) \exp[- F (t)] + f \exp[- F(-t)]\Big\}        \label{27}
    \end{align}
  with $ \tilde E_1 $ being the on-site energy for the side site, which is close to the bridge site energy $E_1, \  \exp [-F (t)] $ being a  dynamic bath correlation function, and $ f $ taking on values $1$ and $0 $ when the attached site is occupied and empty, respectively.

Characterizing the phonon bath with a continuous spectral density $ J (\omega) $ given by Eq. (\ref{17})  one may write out the following expressions for the functions $ F (t) $ and $\delta: $ 
    \be  
  F (t) = \int_0^\infty \frac{d\omega}{\omega^2} J(\omega)
\left[1 - e^{- i\omega t} + \frac{2[1 - \cos (\omega t)] }{ \exp(\hbar \omega/kT)-1}\right],
                            \label{28}                          \ee 
      \be
 \delta = \int_0^\infty \frac{d\omega}{\omega} J(\omega) = J_0  .    \label{29}
          \ee

Within the short time scale $(\omega_c t \ll 1) $ the function $ F(t) $ could be presented in the form \cite{43}:
   \be
F(t) = \frac{J_0}{\hbar\omega_c} \left \{\frac{1}{2}\ln [1 + (\omega_c t)^2]
+ i\arctan (\omega_c t) + K (t) \right \}                                           \label{30}
  \ee
  where
   \be 
 K(t) = (kT)^2 t^2 \zeta \left(2; \frac{kT}{\hbar \omega_c} + 1\right) .         \label{31}
  \ee
   Here, $ \zeta\ \big(2; kT/\hbar \omega_c + 1\big) $ is the Riemann $ \zeta $ function. The asymptotic expression for $ K(t)$ depends on the relation between two parameters, namely, the temperature $ T $ and the cut-off frequency  $ \omega_c $  of the phonon bath.  Assuming $ kT \gg \hbar \omega_c $ 
   \be 
 K(t) \approx \frac{kT}{\hbar \omega_c} (\omega_c t)^2 .      \label{32}
   \ee
 In the opposite limit $\hbar \omega_c \gg kT $ we obtain:
  \be
  K(t) \approx \frac{\pi^2}{6} \Big(\frac{kT t}{\hbar}\Big)^2 \label{33}
  \ee
 Also, we may roughly estimate $K(t)$ within the intermediate range. Taking $ kT \approx \hbar\omega_c  $ we arrive at the approximation $K(t) \approx a^2 (kT t/\hbar)^2 $ where $ a^2 $ is a dimensionless constant of the order of unity.    Correspondingly, within the short time scale we can omit the first term in the expression (\ref{30}), and we get:
  \be 
  F(t) \approx \frac{J_0}{\hbar\omega_c} \big\{i\omega_c t + K(t)\big\}        \label{34}
  \ee
 where $K(t) $ is given by either Eq. (\ref{32}) or Eq. (\ref{33}) depending on the relation between $ \hbar\omega_c $ and $ kT .$

Within the long time scale  $ \omega_c t \gg1, $ and provided that temperatures are not very low $(kT \gg \hbar\omega_c),$  we may present the function $ K(t) $ as:
   \be 
  K(t) = \frac{2 kT t}{\hbar} \int_0^\infty  \frac{dz}{z^2} (1-\cos z)e^{-z/\omega_c t} \approx \frac{\pi kT t}{\hbar}.                           \label{35}                  \ee
  Now, the term $K(t) $ is the greatest addend in the expression for $ F(t),$ so the latter could be approximated as:  $F(t)\approx \pi kT t J_0/\omega_c\hbar^2. $ The same approximation holds within the low temperature limit when $\omega_c\gg kT/\hbar \gg t^{-1}.$

Using the asymptotic expression (\ref{34}), we may calculate the contribution to $ P(E) $ coming from the short time scale $(\omega_c t \ll 1).$ It has the form:
  \begin{align} 
 P_1(E) =& - \frac{i}{2}\sqrt{\frac{\pi}{J_0 kT}} \exp \left 
[-\frac{(E - \tilde E_1)^2}{4 J_0 kT} \right]
  \nn\\ & \times \left 
\{1+ \Phi \left[\frac{i(E - \tilde E_1)}{2\sqrt{J_0 kT}} \right] \right \}       \label{36}
  \end{align}
  where $ \Phi (z) $ is the  probability integral. When both $\hbar \omega_c $ and $ kT $ have the same order of magnitude the expression for $P(E) $ still holds the form (\ref{36}). At $kT \ll \hbar \omega_c, $ the temperature $  kT $ in the expression (\ref{36}) is to be replaced by $ \hbar \omega_c. $  We remark that under the assumption $ kT \gg \hbar \omega_c $ the function $ P_1(E) $ does not depend on the cut-off frequency $\omega_c, $ whereas at $\hbar \omega_c \gg kT $ it does not depend on temperature.  The long time $(\omega_c t \gg 1)$ contribution to $ P(E) $ could be similarly estimated as follows:
   \be 
  P_2(E) = \frac{1}{E-\tilde E_1 + J_0 + i\pi J_0 kT /\hbar \omega_c}.   \label{37}
  \ee
  Comparing these expressions (\ref{36}) and (\ref{37}) we see that the ratio of the peak values of $ P_2(E) $ and $ P_1(E) $ is of the order of $(\hbar^2 \omega_c^2/J_0 kT)^{1/2}.$ Therefore, the term $P_1(E) $ predominates over $P_2(E)$ when the temperatures are moderately high $(\hbar \omega_c < kT)$ and the electron-phonon interaction is not too weak $ J_0/\hbar \omega_c \sim1. $ Usually, experiments on the electrical characterization of conducting polymer nanofibers are carried out at $T\sim 100\div300K,$ so in further analysis we assume that $(\hbar^2 \omega_c^2/J_0 kT)^{1/2}\ll 1,$ and  the term $P_2(E) $  could be omitted. As shown in the Fig. 6,  the imaginary part of $ P(E) $ exhibits a dip around $ E = \tilde E_1 $ and the width of the latter is determined by the product of the temperature $ kT $ (or $ \hbar\omega_c) $ and the constant $ J_0 $ characterizing
the strength of the electron-phonon interaction. When either factor increases, the dip becomes broader and its magnitude reduces.

The presence of the term $ w^2P(E) $ gives rise to very significant changes in the behavior of the Green's function given by the Eq. (\ref{26}). Using the flat band approximation for the self-energy corrections  and disregarding for a while all imaginary terms in the Eq. (\ref{26}), we find that two extra poles of the Green's function emerge.  Assuming $ E_1 = \tilde E_1 = 0 $ and $kT \gg \hbar \omega_c, $ these poles are situated at:
  \be 
 E = \pm 2 \sqrt{J_0 kT |\ln(2J_0 kT/w^2)|}.    \label{38}
  \ee
  The poles correspond to extra electron states  which appear due to electrons coupling to the thermal phonons. These new states are revealed in the structure of the electron transmission $ T(E) $ given by Eq. (\ref{8}).
   The structure of $T(E) $ is shown in the Fig. 6. Two peaks in the transmission are associated with the phonon-induced electronic states. Their positions and heights depend on the temperature  and on the coupling strengths $   J_0, $ and  $ w. $ The important feature in the electron transmission is the absence of the peak associated with the resonance state between the grains (the bridge site) itself. This happens due to the strong suppression of the latter by the effects of the environment. Technically, this peak is damped for it is located at $ E = 0 $ where the imaginary part of $ P(E) $ reachs its maximum in magnitude.  To provide the damping of the original resonance the contribution from the environment (including the side chain attached to the bridge) to the Green's function (\ref{26}) must exceed the terms $\Delta_{1,2} $ describing the effect of the grains. This occurs when the inequality
   \be 
  \Delta_{1,2} <w^2/\sqrt{J_0 kT}        \label{39} 
    \ee
  is satisfied. When the coupling of the bridge to the attached side site is weak, the influence of the environment slackens and the original peak associated with the bridge at $ E= E_1 $  may emerge. At the same time the features originating from the phonon-induced states become small compared to this peak.

\begin{figure}[t] \begin{center}  
 \includegraphics[width=12cm,height=10cm]{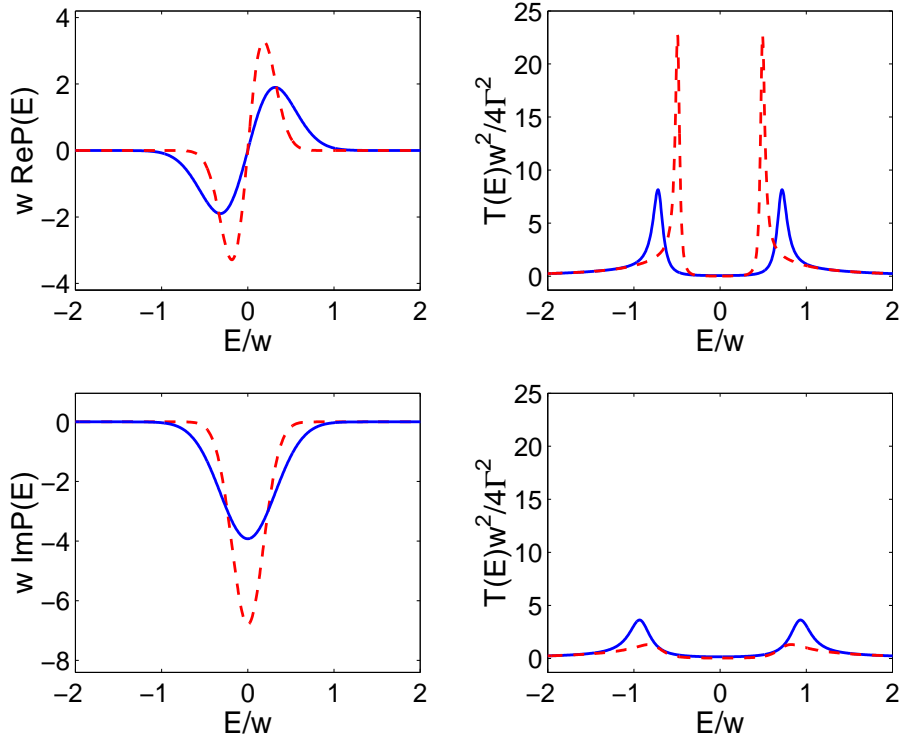}
\caption{(Color online) Left panels: Energy dependence of the real (top panel) and imaginary (bottom panel) parts of $ P(E) .$ The curves are plotted at $ J_0 =20 meV, \ kT \gg \hbar \omega_c. $ Right panels:   The renormalized electron transmission function vs energy.
The constant $ J_0 $ equals $ 20 meV $ (top panel) and $ 50 meV $
(bottom panel). All curves are plotted assuming  $w = 100 meV,\ E_1 = \tilde E_1 = 0,\  T = 100 K $ (dashed lines) and $ T = 300 K $ (solid lines).    } 
\label{rateI}
\end{center}
\end{figure}

So, the effects of the phonons may lead to the
damping of the original resonance state for the electron tunneling between the metallic islands in the polymer fiber. Instead, two phonon-nduced states appear to serve as  intermediate states for the electron transport \cite{44}.  Environmental induced electron states were discussed in the theory of electron conductance  in molecules Refs. \cite{19,45,46}. For instance, it was shown that low biased current-voltage characteristics for molecular junctions with DNA linkers may be noticeably changed due to the occurrence of the phonon-induced electron states similar to these discussed in the present Section. 


As we discussed before, one may treat a nanofiber as a set of parallel working channels, any single channel being a sequence of grains connected with the resonance polymeric chains.  Accordingly, the voltage $ V$ applied across the whole fiber is  distributed among sequental pairs of metallic-like islands included in a single channel. So, in practical nanofibers the voltage $ \Delta V$ applied across two adjacent grains appears to be much smaller than $ V\ (\Delta V/V \sim 10^{-1}-10^{-3}). $   Experiments on the electrical characterization of the polymer fibers are usually carried out at moderately high temperatures $(T \div 300 K)$, so it seems likely that $ kT > \hbar \omega_c. $ Assuming that $ w \sim 100 meV, $ and $ J_0 \sim 20\div 50 meV $ we estimate the separation between the phonon-induced peaks in the electron transmission as $ 120\div 170 meV $ (see Fig. 6). This estimate is close to $ e \Delta V $ when $ V $ takes on values up to $2 \div 3 $ volts. So, the phonon-induced peaks in the electron transmission determine the shape of the current-voltage curves even at reasonably high values of the bias voltage applied across the fiber.
 The resulting current-voltage characteristics are shown in the Fig. 7 (left and middle panels).  The $ I-V $ curves exhibit a nonlinear shape even at room temperature despite the fact that the original state for the resonance tunneling is completely suppressed. This occurs because the intergrain transport is supported by new phonon-induced electron states.

 It is  worthwhile to discuss the temperature dependence of the peak value of the electron transmission which follows from the present results. Using the expression (\ref{26}) for the Green's function and the expression (\ref{36}) for $ P(E) $ we may show that at low temperatures the transmission accepts small values, and exhibits rather weak temperature dependence. At higher temperatures $(T\sim 100K) $ the transmission increases fast as the temperature rises and then it reduces as the temperature further increases. The peak in the transmission is associated with the most favorable conditions for the environment induced states to exist when all remaining parameters (such as $J_0 $ and $w)$ are fixed. At high temperatures the peaks associated with the environment induced states are washed out, as usual. 

\begin{figure}[t]  
\begin{center}   
\includegraphics[width=6cm,height=17.2cm,angle=-90]{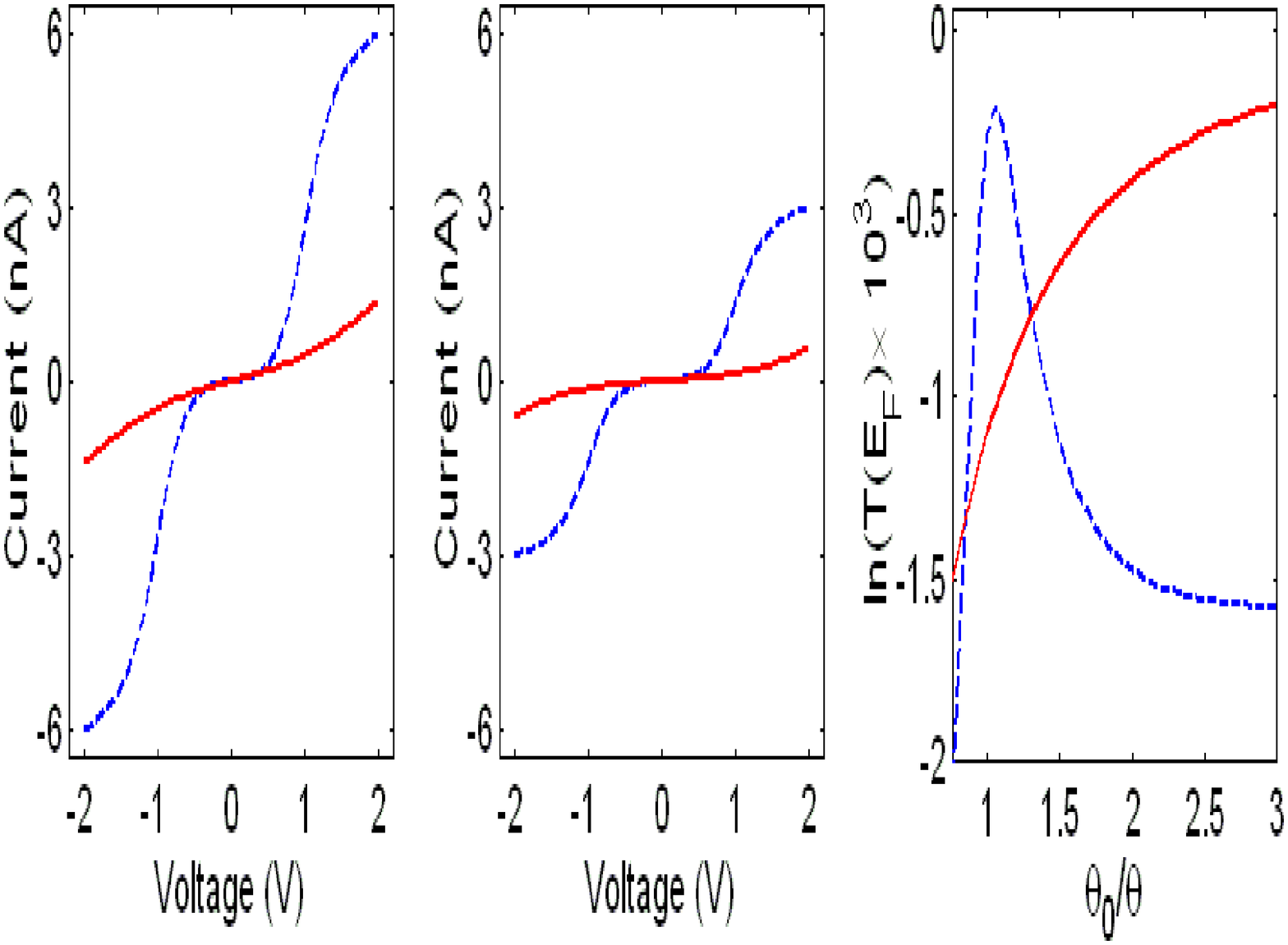}
\caption{
The current-voltage characteristics (nA--V) plotted for $ n =10,\ \  \Delta_1 = \Delta_2 = 0.5 meV,\  \  w = 100 meV,\ \ J_0 = 20 meV $ (left panel), and
$ J_0 = 50 meV $ (midle panel) at $ T = 100  K $ (dash lines) and $ T = 300K $
(solid lines). Right panel: Arrhenius plot of the peak value of the electron transmission function for $ J_0=20  meV,\  w=100  meV,\  \Delta_1=\Delta_2= 0.5  meV,$ and $T_0=100  K .$  The dashed line is plotted assuming the indirect coupling of the bridge to the phonon bath. The solid line is plotted assuming that the bridge is directly coupled to the phonons.
}
\label{rateI}
\end{center}
\end{figure}

We may compare this result with the temperature dependence of the electron transmission function occurring when the bridge between two adjacent grains is directly coupled to the phonon bath. In this case the transmission peak value may be presented in the form determined by Eqs. (\ref{11}), (\ref{24}):
   \be 
  T(E) = 1 - \frac{\rho^4}{2(1+ \sqrt{1+\rho^2})^2}.    \label{40}
  \ee
  The temperature dependencies are shown in the right panel of the Fig. 7. Both curves are plotted at the same value of the electron-phonon coupling strength $ J_0. $ Comparing them we conclude that at higher temperatures the dependencies significantly differ. While the temperature rises, we observe a peak in the electron transmission assuming the indirect coupling of the bridge to the phonon bath, and we see the transmission to monotonically decrease when we consider the bridge directly coupled to the bath. Correspondingly, we may expect qualitative diversities  in the temperature dependencies of the current, as well. These diversities originate from the difference in the effects of environment on the intergrain electron transport in the cases of direct and indirect coupling of the bridge site to the phonon bath. When the bridge is directly coupled to the bath, the stochastic motions in the environment only cause washing out of the peak in the electron transmission, and the higher is the temperature the less distinguishable is the peak. However, when the bridge is screened from the direct coupling with the phonons due to the presence of the nearby sites, the stochastic nuclear motions in the medium between the grains (especially those in the resonance chain) may take a very different part in the electron transport in conducting polymers at moderately low and room temperatures. Due to their influence, the original intermediate state for the resonance tunneling may be completely suppresed but new phonon-induced states may appear to support the electron transport between the metallic-like domains in conducting  polymer nanofibers.


\subsection{\normalsize  V. Conclusion}

Studies of the electron transport in conducting polymers are not completed so far. Several mechanisms are known to control  the charge transport in these highly disordered
and inhomogeneous materials,  and their relative significance could vary depending on both specific intrinsic characteristics of a particular material (such as crystallization rate and electron-electron and electron-phonon coupling strengths)  and  external factors
such as temperature. Various conduction mechanisms give rise to various temperature dependencies of the electric current and conductance, which could be observed in polymer
nanofibers/nanotubes being in conducting state. In the present work we 
aimed at finding out the character of temperature dependencies of both current and conductance provided by
specific transport mechanism, namely, resonance tunneling of electrons.

Accordingly, we treated a conducting polymer as a kind of granular metal, and we assumed that the intergrain conduction occurred due to the electron tunneling between the metallic-like grains through the intermediate state. This scenario for the intergrain electron transport strongly resembles electron transport through molecules/quantum dots attached to the conducting leads. In the considered case the metallic-like islands work as the leads and the intermediate state in between acts as a single level quantum dot/molecular bridge. Basing on this similarity we did apply the well-known Landauer formula to compute the intergrain electron current. This bringed results which agreed with experiments on electrical characterization of doped polyaniline-polyethylene oxide nanofibers.

There are solid grounds to expect significant dissipative effects in the intergrain transport at moderately high temperatures. To take into account the effect of temperature we did represent the thermal environment  stochastic nuclear motions as a phonon bath, and we introduced the coupling of the intermediate site to the thermal phonons.  As shown in the previous studies of the electron transport through molecules, various dissipative effects may occur depending on  characteristic features in the interaction of a propagating electron with the environment. Among these features we  singled out the character of the electron coupling to the dissipative reservoir (phonon bath) as a very significant factor. It is likely that in practical conducting polymers both direct and indirect coupling of the intermediate state (the bridge) to the environment may occur.

We did analyze temperature dependencies of transport characteristics for both scenarios and  we showed that these dependencies  differ. Also, we showed that in general, resonance electron tunneling between the grains results in the temperature dependencies of transport characteristics, which differ from those obtained for other conduction mechanisms such as phonon-assisted hopping between  localized states. Being observed in experiments on realistic polymer nanofibers, the predicted dependencies would give grounds to suggest the electron tunneling to predominate in the intergrain electron transport in these particular nanofibers. We believe the present studies to contribute to better understanding of electron transport mechanisms in conducting polymers.
\vspace{2mm}

{\bf Acknowledgments:} Author  thanks  G. M. Zimbovsky for help with the 
manuscript. This work was supported  by NSF-DMR-PREM 0353730.



\begin{thebibliography}{99}
\bibitem{39} The matter is thoroughly discussed by Kaiser (see \cite{2} and
references therein).
\end{thebibliography}
\end{document}